\documentclass[twocolumn]{aastex7}

\usepackage{amsmath}  
\usepackage{amsfonts}
\usepackage{graphicx}

\providecommand{\ion}[2]{\mathrm{#1}\,\textsc{#2}}

\begin{document}

\title{Balmer Break Measurements Trace Departures from the Star-Forming Main Sequence at \mbox{\boldmath 1\,$<$\,$z$\,$<$\,4} with JWST/NIRSpec}
\author[0009-0006-5305-8363]{Suprabhas Narisetty}
\affiliation{Department of Astronomy, University of Illinois Urbana-Champaign, 1002 W. Green Street, Urbana, IL 61801, USA}
\affiliation{Steward Observatory, University of Arizona, 933 North Cherry Avenue, Tucson, AZ 85721, USA}
\email[show]{ssn13@illinois.edu}

\author[0000-0003-3307-7525]{Yongda Zhu}
\affiliation{Steward Observatory, University of Arizona, 933 North Cherry Avenue, Tucson, AZ 85721, USA}
\email{yongdaz@arizona.edu}

\author[0000-0003-1344-9475]{Eiichi Egami}
\affiliation{Steward Observatory, University of Arizona, 933 North Cherry Avenue, Tucson, AZ 85721, USA}
\email{egami@arizona.edu}

\author[0000-0002-6719-380X]{Stefano Carniani}
\affiliation{Scuola Normale Superiore, Piazza dei Cavalieri 7, I-56126 Pisa, Italy}
\email{stefano.carniani@sns.it}

\author[0000-0003-2388-8172]{Francesco D'Eugenio}
\affiliation{Kavli Institute for Cosmology, University of Cambridge, Madingley Road, Cambridge, CB3 0HA, UK}
\affiliation{Cavendish Laboratory, University of Cambridge, 19 JJ Thomson Avenue, Cambridge, CB3 0HE, UK}
\email{fd391@cam.ac.uk}

\author[0000-0002-4781-9078]{Christa DeCoursey}
\affiliation{Steward Observatory, University of Arizona, 933 North Cherry Avenue, Tucson, AZ 85721, USA}
\email{cndecoursey@arizona.edu}

\author[0000-0001-9262-9997]{Christopher N. A. Willmer}
\affiliation{Steward Observatory, University of Arizona, 933 North Cherry Avenue, Tucson, AZ 85721, USA}
\email{cnaw@arizona.edu}

\begin{abstract}
We present JWST/NIRSpec PRISM observations of galaxies at $1<z<4$ to study the relation between Balmer-break strength ($D_{\rm Balmer}$) and recent star formation. Using NIRSpec spectra from the JADES Transient Survey in GOODS-S, we measure $D_{\rm Balmer}$ and hydrogen recombination-line equivalent widths, and infer galaxy properties through spectral energy distribution (SED) fitting of the accompanying photometry. We find a clear anti-correlation between $D_{\rm Balmer}$ and the rest-frame equivalent width of H$\alpha$+[\ion{N}{2}], such that galaxies with stronger breaks show weaker nebular emission and lower recent star formation activity. $D_{\rm Balmer}$ also correlates with mass-weighted stellar age, supporting its use as an empirical tracer of evolved stellar populations at these redshifts. Star formation rates inferred from H$\alpha$, Pa$\gamma$, and SED fitting are broadly consistent, with scatter likely driven in part by dust attenuation, line-measurement uncertainties, and the different timescales probed by each indicator. Galaxies with stronger $D_{\rm Balmer}$ also tend to lie further below the star-forming main sequence at fixed stellar mass, indicating that $D_{\rm Balmer}$ is linked to recent suppression of star formation. In contrast, we find only a weak relation between $D_{\rm Balmer}$ and SED-derived dust attenuation, suggesting that dust is not the main driver of the observed trends. These results show that $D_{\rm Balmer}$, combined with recombination-line measurements, is a useful empirical diagnostic of galaxies transitioning away from the star-forming population at cosmic noon.
\end{abstract}

\section{Introduction}

Quenching at cosmic noon redshifts ($1\lesssim z \lesssim 4$) remains a relatively poorly understood process. Understanding how and when galaxies reduce their star formation during this epoch is critical for constraining galaxy growth. This period corresponds to the peak of the cosmic star formation rate density \citep{madau_cosmic_2014}, and yet a substantial population of galaxies is already transitioning toward quiescence \citep{bevacqua_tp-agb_2025}. The existence of quiescent galaxies at $z > 2$ implies that some galaxies undergo a rapid decline in star formation activity after an earlier phase of stellar mass growth. For example, \citet{forrest_extremely_2020} found that a massive quiescent galaxy at $z = 3.493$ experienced a burst of star formation lasting approximately 0.5~Gyr, followed by an abrupt truncation.

This raises a fundamental question: what mechanisms drive this rapid decline in star formation in these galaxies? A wide range of processes has been proposed, including gas depletion, active galactic nuclei (AGN) feedback, and dynamical effects associated with mergers. In gas-rich systems, mergers may trigger starbursts that either exhaust or expel the gas reservoir, leading to rapid quenching on timescales of hundreds of Myr \citep{mihos_gasdynamics_1996,hopkins_cosmological_2008}. More broadly, galaxy-scale outflows and feedback are expected to play a major role in regulating star formation and driving galaxy evolution \citep{veilleux_galactic_2005,somerville_physical_2015,naab_theoretical_2017}. Recent observations have also linked powerful outflows to feedback in high-redshift galaxies and AGN hosts, including with James Webb Space Telescope \citep{perrotta_kinematics_2023,zhu_systematic_2025,zhu_smiles_2025}. Other studies suggest that AGN feedback regulates star formation in the most massive galaxies, and JWST has also enabled improved characterization of the role of AGN in massive galaxies at high redshift \citep{schawinski_observational_2007,springel_black_2005,rieke_confirming_2025,lyu_active_2024}. 

In this work, we focus on lower-mass galaxies with $\log_{10}(M_\star/M_\odot)\lesssim9.75$. In this mass regime, recent suppression of star formation must be considered against bursty star formation histories. The THESAN-ZOOM simulations illustrate how gas inflows, ejection and re-accretion can produce alternating periods of increased and suppressed star formation \citep{mcclymont_rinse_repeat_2025}. As a result, we should examine how Balmer break strength and nebular emission trace recent star formation suppression in the lower-mass regime. 

Post-starburst galaxies offer a valuable opportunity to trace this transition. These systems are characterized by strong Balmer absorption and weak nebular emission, indicating that star formation has recently declined or ended. They can therefore be used to constrain quenching timescales and to probe the transition phase in galaxy evolution. At $z > 2$, a large fraction of quiescent galaxies appears to have undergone rapid quenching, consistent with expectations for a post-starburst phase \citep{french_evolution_2021,ji_jems_2024,park_bluejay_2024}. Our work considers a broader galaxy population, without requiring a post-starburst qualification. We use the Balmer break strength and nebular emission to study recent star formation suppression across this population. 

Effective spectral diagnostics, such as the strength of the Balmer break and the H$\alpha$ emission line, are powerful tools for studying stellar population ages and recent star formation histories. In a galaxy's stellar population, the Balmer break is most pronounced when A-type stars dominate the spectrum. Since such stars have main-sequence lifetimes of $\sim 0.1$--1~Gyr, a strong Balmer break indicates a substantial population of intermediate-age stars \citep[e.g.][]{deugenio_typical_2020}. In practice, the presence of a strong Balmer break implies that recent star formation has declined enough for the galaxy's spectrum to be dominated by older stars instead of hot O- and B-type stars. Balmer-break strength is therefore an effective indicator of recent star formation history, as it peaks in the aftermath of a burst. Meanwhile, H$\alpha$ emission is sensitive to ongoing star formation because it traces the ionizing photons from hot, short-lived O/B stars. Together, these two indicators allow us to distinguish actively star-forming galaxies from those that have recently been quenched. The Balmer break at 3646~\AA{} is distinct from the
4000-\AA{} break, which arises primarily from the accumulation
of absorption lines below approximately 4000~\AA{} and becomes
more prominent in older, metal-rich stellar populations.
The classical $D_{4000}$ \citep{Bruzual1983} and
$D_{n,4000}$ \citep{balogh_differential_1999} indices measure
this latter feature using continuum windows entirely redward
of the Balmer edge. Here, we instead measure the continuum
contrast across the Balmer edge using our $D_{\rm Balmer}$ index.

At cosmic noon, however, measuring the Balmer break spectroscopically has been difficult because the relevant rest-frame optical features are shifted into the near-infrared. While recent studies have focused on the evolution of the Balmer break at high redshift ($z > 3.5$; \citealt{Kuruvanthodi2026BalmerBreak}), understanding its evolution during cosmic noon is equally important and would benefit from a uniform sample of galaxies observed with JWST/NIRSpec \citep{jakobsen_near-infrared_2022}. JWST/NIRSpec \citep{jakobsen_near-infrared_2022} now makes such measurements possible for large samples of galaxies at these redshifts, owing to its sensitivity and wavelength coverage. Recent surveys and public data sets, such as JADES \citep{eisenstein_jades_2023, tacchella_jades_2023, rieke_marcia_data_2024,bunker_jades_2024,deugenio_jades_2025,curtis-lake_jades_2025,scholtz_jades_2025}, CEERS \citep{finkelstein_ceers_2023}, UNCOVER \citep{bezanson_jwst_2024}, RUBIES \citep{de_graaff_rubies_2025}, and SMILES \citep{alberts_smiles_2024, rieke_smiles_2024, zhu_smiles_2025}, have demonstrated the power of JWST spectroscopy for studying galaxy stellar populations, nebular emission, and quenching-related diagnostics at cosmic noon and beyond.

In this work, we analyze galaxies at $1\lesssim z \lesssim 4$ with JWST/NIRSpec \citep{jakobsen_near-infrared_2022} PRISM spectra to quantify how Balmer-break strength and H$\alpha$ emission trace recently suppressed star formation. We explore how the break strength correlates with H$\alpha$ equivalent width, dust attenuation, star formation rate, and offset from the star-forming main sequence.

This paper is organized as follows. Section \ref{sec:data} describes the data and sample. Section \ref{sec:method} outlines our spectral measurements and SED fitting methods. Section \ref{sec:results} presents the results, and Section \ref{sec:discussion} discusses their implications. We summarize the findings in Section \ref{sec:summary}. Throughout this paper, we assume a flat $\Lambda$CDM cosmology with $\Omega_{m} = 0.315$, $\Omega_{\Lambda} = 0.685$, and $H_{0} = 67.4\,\mathrm{km\,s^{-1}\,Mpc^{-1}}$ for luminosity calculations \citep{planck_collaboration_planck_2018}.

\section{Data and Sample} \label{sec:data}

The parent sample is drawn from the JWST Advanced Deep Extragalactic Survey (JADES) in the GOODS-South field \citep{eisenstein_jades_2023,rieke_jades_2023-1}. The spectra were obtained with the low-resolution NIRSpec PRISM mode, which provides continuous wavelength coverage from 0.6 to 5.3~$\mu$m at $R\sim100$. This wavelength range covers the Balmer and the 4000~\AA{} continuum break and the main rest-frame optical recombination lines for galaxies at $1<z<4$.

Data reduction follows the JADES NIRSpec Data Release 3 pipeline \citep{deugenio_jades_2025}, which uses custom tools developed by the GTO team \citep{boker_near-infrared_2022}. The galaxies analyzed here come from NIRSpec PRISM observations obtained as part of the JADES Transient Survey \citep[][Program ID 6541]{decoursey_jades_2025}. Spectroscopic
observations were obtained with the NIRSpec Micro-Shutter Assembly
(MSA; \citealt{Ferruit2022}), obtained from Observation 1. The
MSA setup consisted of three dithered configurations, each observed
in three nodded exposures. Each exposure used 19 groups and 2 integrations, lasting 2801 s each. The total time on source ranges
from $2801\times3=8403$~s (for targets observed in only one MSA
configuration) to $2801\times9=25209$~s (for targets observed in
all three MSA configurations). The final reduced list includes 149
galaxy spectra spanning redshifts up to $z\sim6$.

Spectroscopic redshifts were first estimated visually based on JADES photometric-redshift priors \citep{hainline_cosmos_2024}, similar to the procedure used by the JADES and SMILES teams \citep{zhu_smiles_2025}. We then fit the spectra with {\tt GELATO} \citep[Gaussian Emission Line Analysis Tool;][]{hviding_theskyentistgelato_2022} and confirmed the redshifts through the alignment of detected emission lines, continuum breaks, and the overall spectral shape. For galaxies with noisy spectra around prominent emission lines such as H$\alpha$, H$\beta$, and [O~\textsc{iii}], the photometric redshift was used as an initial prior, but the final adopted redshift required consistency with the observed spectrum.

From this parent sample of 149 galaxies, we selected galaxies at $1<z<4$ with spectral coverage of the continuum windows used to measure $D_{\rm Balmer}$, the ratio between the blue and red sides of the continuum break. This redshift range ensures that the Balmer-break region and the main rest-frame optical recombination lines fall within the NIRSpec/PRISM wavelength coverage. The resulting sample includes galaxies with a range of Balmer-break strengths, allowing us to test how the continuum break relates to stellar population age, emission-line equivalent width, dust attenuation, and star formation activity. The detailed quality cuts for the $D_{\rm Balmer}$, emission-line, and SED-based measurements are described in the relevant Methods and Results subsections.

The parent sample contains 149 galaxy spectra. The
main analysis sample includes 57 unique galaxies for which we can reliably measure $D_{\rm Balmer}$. The mass-weighted age analysis uses that sample, while the stellar dust-attenuation analysis includes 54.
The equivalent-width--$D_{\rm Balmer}$ analysis includes
29 galaxies, and the SED-based main-sequence and
main-sequence-offset analyses include 38. The two panels
comparing our sample with \citet{kriek_h_2011} contain
41 galaxies for the break-strength comparison and
38 for the sSFR comparison. Sample membership depends
on the measurements required for each analysis; the
corresponding selection criteria are described below.

\begin{figure*}[!ht]
    \centering
    \includegraphics[width=1\linewidth]{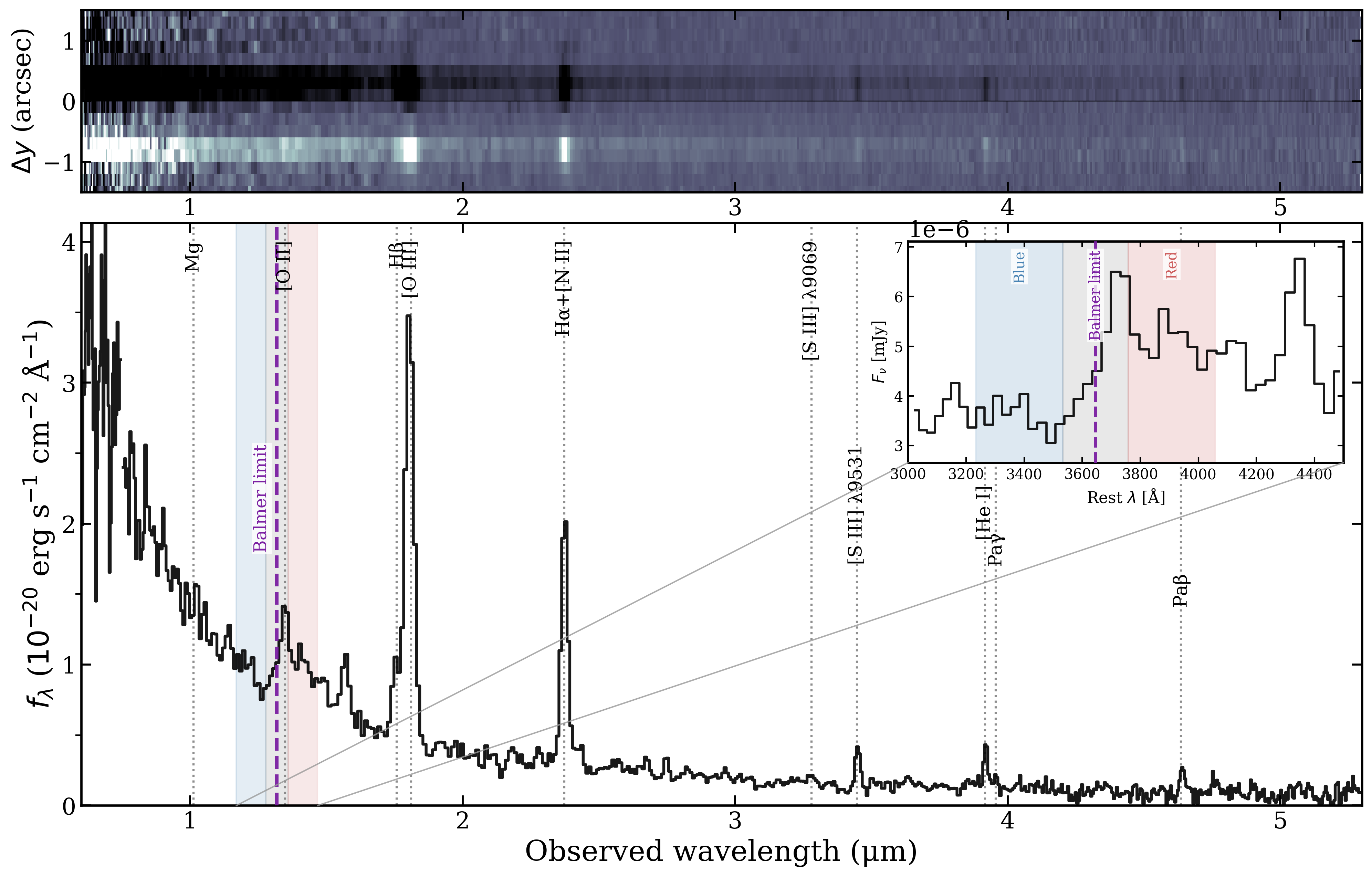}
    \caption{Example JWST/NIRSpec PRISM spectrum used for the Balmer-break
    measurement. The top panel shows the two-dimensional spectrum,
    while the main bottom panel shows the one-dimensional
    observed-frame spectrum in $F_\lambda$. The observed spectrum
    is shown in black. The dashed purple line marks the Balmer
    limit at a rest-frame wavelength of 3646~\AA{}. The lightly
    shaded blue and red regions mark the blue- and red-side
    continuum windows, respectively, while the gray region shows
    the excluded part of the spectrum. The main emission lines
    are also indicated on the spectrum. The inset shows the
    rest-frame spectrum around the break region in $F_\nu$,
    with labels identifying the blue and red windows and the
    Balmer limit. Galaxy 097858
    (RA = 53.116455$^\circ$, DEC = $-27.8198357^\circ$)
    has a spectroscopic redshift of $z=2.6165$ and a measured break strength of
    $D_{\rm Balmer}=1.423^{+0.091}_{-0.080}$. }
    \label{fig:balmerbreak}
\end{figure*}

\section{Methods}\label{sec:method}

This section describes the methodology used to derive the strength of the continuum breaks, emission-line properties and physical galaxy parameters from the JWST/NIRSpec PRISM spectra and the corresponding JADES photometry. The PRISM data has low resolution, so our analysis was designed to obtain physically meaningful and consistent results while accounting for limitations such as blended spectral features. We adopted a modified break definition from the norm \citep{balogh_differential_1999, hamilton_spectral_1985}, designed to account for the low resolution. We tailored the emission-line fitting choices to the spectra and used broadband SED fitting to derive key properties. Together, these methods provide the framework for comparing recent star formation, continuum features and galaxy properties for the sample.

\subsection{\texorpdfstring{$D_{\rm Balmer}$}{D Balmer} Break Determination}

We quantify the Balmer-break strength using $D_{\rm Balmer}$,
defined as the ratio of the median $F_\nu$ flux density in a
red-side continuum window to that in a blue-side window
bracketing the 3646-\AA{} Balmer edge. These windows sample
the broader continuum contrast across the edge rather than
the discontinuity alone.

The D$_{\rm Balmer}$ index was measured as the ratio between the median flux density of the two rest-frame windows: red and blue. Another modification was implemented at this stage; due to the low resolution ($R\sim 100$) of the PRISM spectra, the initial continuum window width of 200 \AA{} increased by a factor of 1.5 to 300 \AA{}. Similarly, the gap between the two windows, initially at a width of 150 \AA{}, increased by the same factor, resulting in a 225 \AA{} gap. The purpose of this gap is to encapsulate the rapid change in the $F_\nu$ value. Thus, the red window is between 3758.5 \AA and 4058.5 \AA, and the blue window is between 3233.5 \AA and 3533.5 \AA. We introduce this break to avoid biasing the flux from the rapid decrease in the spectra. These modifications help reduce the impact of limited resolution
and potentially high spectral noise. To ensure reliable break measurements, we required that the fractional uncertainty of $D_{\rm Balmer}$ to be $\leq 20\%$, corresponding to a signal-to-noise ratio (SNR) of at least 5. We also excluded galaxies with measured break strengths outside $1.0<D_{\rm Balmer}<2.0$. Visual inspections of the spectra indicated that these extreme measurements do not reflect genuine continuum spectra.

We compute the Balmer break strength in $F_\nu$ units as
\begin{equation}
D_{\rm Balmer} = \frac{\langle F_{\nu,\mathrm{red}} \rangle}{\langle F_{\nu,\mathrm{blue}} \rangle}\,.
\end{equation}
Figure \ref{fig:balmerbreak} shows an example of the measurement.

\subsection{Emission Line Fitting}

We fit hydrogen recombination lines from the Balmer and Paschen series and the nebular lines of [\ion{Ne}{5}], [\ion{Ne}{3}], [\ion{O}{3}], [\ion{N}{2}], [\ion{S}{2}], [\ion{O}{1}], [\ion{O}{2}], and [\ion{S}{3}] using {\tt GELATO} \citep[Gaussian Emission Line Analysis Tool;][]{hviding_theskyentistgelato_2022}. The code performs non-linear Gaussian modeling and bootstrapped uncertainty calculation. In our fitting configuration, the Balmer and Paschen series were tied in wavelength only. The [\ion{O}{1}], [\ion{O}{2}], and [\ion{S}{3}] group was tied in wavelength and velocity dispersion, while the [\ion{Ne}{5}], [\ion{Ne}{3}], [\ion{O}{3}], [\ion{N}{2}], and [\ion{S}{2}] lines were left untied. These groups specify fitting constraints rather than classifications of the ionizing source.

The {\tt GELATO} configuration used for this study followed the tool's default settings \footnote{\url{https://github.com/TheSkyentist/GELATO}}, with the exception of two key fit parameters: the velocity width around each emission line (\texttt{LineRegion}) was set to 4000~km\,s$^{-1}$, and the region excluded for continuum fitting around each emission line (\texttt{ContinuumRegion}) was set to 15000~km\,s$^{-1}$. These choices were adopted to improve fitting stability for the low-resolution PRISM spectra.

The final output of the code included fluxes, equivalent widths, redshifts, and dispersions of emission lines. These measurements were used to compare the spectroscopic, emission line based star formation metrics with continuum-based quantities. Because H$\alpha$ and [N~\textsc{ii}] are blended at PRISM resolution, we use the blended H$\alpha$+[N~\textsc{ii}] equivalent width as the primary emission-line diagnostic in this work.

\subsection{Spectral Energy Distribution Fitting}

Galaxy physical properties and star-formation histories were inferred via SED fitting using the {\tt PROSPECTOR} code \citep{johnson_stellar_2021}. We used Kron-aperture photometry from the JADES DR1 NIRCam GOODS-S Deep photometric catalog \citep{rieke_jades_2023}. {\tt PROSPECTOR} combines FSPS-based stellar population synthesis models \citep{byler_nebular_2017} with nebular emission from previously computed CLOUDY photo-ionization grids \citep{ferland_2013_2013}, so that the continuum and line emission can be modeled together.

Dust attenuation was modeled using the two-component model of \citet{charlot_simple_2000}, where the dust index is allowed to vary between $-1.2$ and $0.4$. The dust attenuation for stars younger than 10 Myr was corrected through a functional dependence on {\tt dust2}, while the {\tt dust1} component was allowed to vary freely with a truncated normal prior centered at 1.0.

We used a non-parametric star formation history made up of seven time bins from 0 to 2000 Myr, with widths of 5 Myr ([0--5] Myr), 5 Myr ([5--10] Myr), 10 Myr ([10--20] Myr), 30 Myr ([20--50] Myr), and 50 Myr ([50--100] Myr). The final two bins were broader, each covering 400 Myr ([100--500] Myr) and 1500 Myr ([500--2000] Myr) to record longer-term bursts or quenching. We adopted the continuity prior of \citet{leja_2019}, assigning the logarithmic SFR ratios between adjacent time bins a Student's $t$-distribution centered at zero, with scale $\sigma=0.3$ and $\nu=2$ degrees of freedom. A \citet{chabrier_galactic_2003} IMF was assumed and the stellar metallicity. gas-phase metallicity, and nebular ionization parameter $\log U$ were defined as free parameters.

\section{Results} \label{sec:results}

\subsection{Sample Properties}

Tables~\ref{tab:measurements} and
\ref{tab:derived_properties} summarize the measured
and inferred properties of the 57 unique galaxies
contributing to our analyses. The sample size varies
between analyses according to the availability and
quality of the required measurements. For poor emission-line detections with $F/\sigma_F<2$, including negative fitted fluxes, we report $2\sigma$ upper limits on the H$\alpha$ and Pa$\gamma$ fluxes and corresponding SFRs, calculated using twice the measured line-flux uncertainty.

\startlongtable
\begin{deluxetable*}{lcccccccc}
\tabletypesize{\scriptsize}
\tablewidth{0pt}
\tablecaption{Catalog information and spectroscopic measurements.\label{tab:measurements}}
\tablehead{
\colhead{JADES ID} & \colhead{RA} & \colhead{DEC} & \colhead{$z_{\rm spec}$} & \colhead{$z_{\rm phot}$} & \colhead{$D_{\rm Balmer}$} & \colhead{EW(H$\alpha$+[N\,II])} & \colhead{$F_{\rm H\alpha}$} & \colhead{$F_{\rm Pa\gamma}$} \\
\colhead{---} & \colhead{deg} & \colhead{deg} & \colhead{---} & \colhead{---} & \colhead{---} & \colhead{\AA} & \colhead{$10^{-18}\,{\rm erg\,s^{-1}\,cm^{-2}}$} & \colhead{$10^{-18}\,{\rm erg\,s^{-1}\,cm^{-2}}$}
}
\startdata
083749 & $53.134359$ & $-27.837081$ & $3.7066$ & $4.22$ & $1.28_{-0.07}^{+0.08}$ & $292.4\pm7.9$ & $1.125\pm0.045$ & $0.263\pm0.059$ \\
085339 & $53.114741$ & $-27.835152$ & $2.8512$ & $2.78$ & $1.46_{-0.07}^{+0.07}$ & $511.9\pm12.7$ & $3.746\pm0.157$ & $0.233\pm0.035$ \\
085407 & $53.141298$ & $-27.835062$ & $2.5746$ & $2.48$ & $1.74_{-0.14}^{+0.14}$ & $-6.6\pm0.1$ & $<0.011$ & $<0.006$ \\
085997 & $53.177799$ & $-27.834180$ & $1.5613$ & $1.44$ & $1.94_{-0.26}^{+0.30}$ & $27.3\pm22.2$ & $<0.176$ & $<0.079$ \\
086866 & $53.129196$ & $-27.833039$ & $1.6359$ & $1.53$ & $1.48_{-0.08}^{+0.09}$ & $-5.1\pm8.3$ & $<0.391$ & $<0.047$ \\
087622 & $53.172022$ & $-27.832056$ & $2.8984$ & $2.90$ & $1.48_{-0.19}^{+0.26}$ & $320.2\pm33.3$ & $0.441\pm0.178$ & $<0.207$ \\
088115 & $53.113160$ & $-27.831317$ & $1.0955$ & $1.09$ & $1.53_{-0.09}^{+0.09}$ & $26.5\pm7.1$ & $0.651\pm0.230$ & $0.178\pm0.052$ \\
089220 & $53.128098$ & $-27.829750$ & $1.3701$ & $1.34$ & $1.47_{-0.12}^{+0.13}$ & $3.3\pm4.1$ & $<2.6e-05$ & $<0.129$ \\
089509 & $53.175828$ & $-27.829319$ & $2.7997$ & $1.68$ & $1.45_{-0.22}^{+0.22}$ & $224.0\pm9.9$ & $0.365\pm0.038$ & $<0.083$ \\
090417 & $53.127627$ & $-27.828017$ & $3.0206$ & $3.04$ & $1.47_{-0.04}^{+0.04}$ & $716.3\pm14.9$ & $7.913\pm0.534$ & $1.371\pm0.047$ \\
090735 & $53.131981$ & $-27.827623$ & $2.6845$ & $2.79$ & $1.42_{-0.04}^{+0.04}$ & $103.9\pm13.2$ & $1.188\pm0.484$ & $0.362\pm0.037$ \\
092192 & $53.118592$ & $-27.825680$ & $1.3583$ & $1.42$ & $1.81_{-0.06}^{+0.07}$ & $17.0\pm2.9$ & $0.521\pm0.025$ & $0.380\pm0.084$ \\
093268 & $53.140616$ & $-27.824459$ & $1.8918$ & $1.90$ & $1.51_{-0.11}^{+0.12}$ & $40.1\pm9.1$ & $0.283\pm0.113$ & $<0.065$ \\
094972 & $53.177461$ & $-27.822786$ & $3.4574$ & $3.47$ & $1.22_{-0.12}^{+0.12}$ & $634.2\pm5.2$ & $1.723\pm0.000$ & $<0.083$ \\
097581 & $53.140822$ & $-27.820095$ & $3.9660$ & $4.15$ & $1.33_{-0.08}^{+0.09}$ & $461.0\pm13.0$ & $1.803\pm0.067$ & \nodata \\
097858 & $53.116455$ & $-27.819836$ & $2.6165$ & $2.63$ & $1.42_{-0.08}^{+0.09}$ & $75.8\pm3.3$ & $<1.7e-06$ & $0.171\pm0.042$ \\
099110 & $53.136840$ & $-27.818394$ & $2.5898$ & $2.65$ & $1.60_{-0.10}^{+0.12}$ & $231.6\pm148.3$ & $<0.274$ & $<0.193$ \\
115313 & $53.136795$ & $-27.794686$ & $4.0206$ & $4.22$ & $1.09_{-0.41}^{+0.69}$ & $881.5\pm92.1$ & $0.605\pm0.056$ & \nodata \\
115803 & $53.177386$ & $-27.793757$ & $2.3111$ & $2.34$ & $1.39_{-0.11}^{+0.13}$ & $45.9\pm8.2$ & $0.062\pm0.027$ & $<0.044$ \\
116235 & $53.132056$ & $-27.793189$ & $2.3308$ & $2.31$ & $1.27_{-0.14}^{+0.16}$ & $243.9\pm18.9$ & $1.033\pm0.095$ & $0.084\pm0.036$ \\
116687 & $53.116504$ & $-27.792407$ & $2.0783$ & $2.00$ & $1.76_{-0.13}^{+0.11}$ & $36.7\pm3.2$ & $0.370\pm0.055$ & $0.089\pm0.042$ \\
121632 & $53.139186$ & $-27.786233$ & $1.9966$ & $1.99$ & $1.47_{-0.11}^{+0.12}$ & $295.4\pm89.7$ & $<2.902$ & $0.349\pm0.067$ \\
123725 & $53.123941$ & $-27.783940$ & $1.0507$ & $1.08$ & $1.67_{-0.20}^{+0.24}$ & $4.7\pm5.6$ & $<0.149$ & $<0.075$ \\
123903 & $53.166388$ & $-27.783647$ & $3.3319$ & $3.36$ & $1.47_{-0.18}^{+0.21}$ & $361.9\pm41.5$ & $0.826\pm0.085$ & $<0.120$ \\
124533 & $53.119596$ & $-27.782806$ & $2.3083$ & $2.31$ & $1.51_{-0.08}^{+0.08}$ & $280.6\pm16.5$ & $0.580\pm0.274$ & $0.194\pm0.029$ \\
125100 & $53.131155$ & $-27.782050$ & $2.1442$ & $2.40$ & $1.34_{-0.10}^{+0.11}$ & $9.2\pm48.8$ & $<0.122$ & $<0.066$ \\
125748 & $53.116198$ & $-27.781094$ & $2.8686$ & $2.91$ & $1.72_{-0.05}^{+0.05}$ & $86.7\pm4.0$ & $<0.311$ & $0.220\pm0.055$ \\
126238 & $53.125176$ & $-27.780388$ & $3.7713$ & $0.68$ & $1.28_{-0.15}^{+0.18}$ & $1214.6\pm37.9$ & $0.966\pm0.093$ & $0.206\pm0.098$ \\
191229 & $53.172124$ & $-27.836383$ & $1.1366$ & $1.21$ & $1.83_{-0.09}^{+0.12}$ & $57.2\pm15.8$ & $0.853\pm0.133$ & $<0.139$ \\
191896 & $53.172746$ & $-27.834389$ & $2.2714$ & $2.48$ & $1.39_{-0.11}^{+0.12}$ & $45.0\pm14.2$ & $<0.107$ & $<0.021$ \\
192046 & $53.134539$ & $-27.833996$ & $1.8200$ & $1.69$ & $1.47_{-0.08}^{+0.08}$ & $91.6\pm20.5$ & $0.476\pm0.190$ & $<0.087$ \\
192594 & $53.113170$ & $-27.832142$ & $1.2065$ & $1.34$ & $1.86_{-0.14}^{+0.16}$ & $28.4\pm24.9$ & $<5.085$ & $<0.433$ \\
193779 & $53.178186$ & $-27.828391$ & $2.8097$ & $2.87$ & $1.48_{-0.02}^{+0.02}$ & $169.6\pm43.5$ & $<3.877$ & $0.287\pm0.047$ \\
194960 & $53.139420$ & $-27.824787$ & $2.8257$ & $2.79$ & $1.68_{-0.05}^{+0.05}$ & $170.3\pm7.4$ & $1.603\pm0.348$ & $0.494\pm0.056$ \\
195115 & $53.127054$ & $-27.824719$ & $3.0287$ & $3.03$ & $1.15_{-0.08}^{+0.08}$ & \nodata & $<3.968$ & $0.650\pm0.132$ \\
195924 & $53.138726$ & $-27.822057$ & $2.5864$ & $2.63$ & $1.63_{-0.08}^{+0.09}$ & $153.5\pm25.6$ & $<0.717$ & $0.066\pm0.023$ \\
196204 & $53.174593$ & $-27.821829$ & $2.6132$ & $2.64$ & $1.12_{-0.09}^{+0.10}$ & $463.4\pm39.2$ & $0.592\pm0.217$ & $0.100\pm0.025$ \\
197390 & $53.139957$ & $-27.817822$ & $2.0146$ & $2.04$ & $1.83_{-0.13}^{+0.16}$ & $64.7\pm6.1$ & $0.951\pm0.114$ & $0.120\pm0.047$ \\
198754 & $53.114827$ & $-27.813230$ & $1.5542$ & $1.56$ & $1.68_{-0.03}^{+0.03}$ & $17.4\pm20.4$ & $<5.883$ & $<0.061$ \\
201169 & $53.163837$ & $-27.804182$ & $2.3088$ & $2.31$ & $1.30_{-0.03}^{+0.03}$ & $120.7\pm194.7$ & $0.422\pm0.155$ & $0.395\pm0.028$ \\
202445 & $53.178890$ & $-27.799634$ & $3.2694$ & $3.31$ & $1.58_{-0.13}^{+0.12}$ & $179.4\pm18.1$ & $0.873\pm0.140$ & $<0.101$ \\
202882 & $53.137655$ & $-27.797698$ & $1.8904$ & $1.85$ & $1.66_{-0.12}^{+0.14}$ & $6.4\pm29.9$ & $<1.703$ & $0.207\pm0.034$ \\
203991 & $53.163616$ & $-27.793602$ & $1.3603$ & $1.34$ & $1.72_{-0.04}^{+0.05}$ & $71.9\pm2.6$ & $1.093\pm0.251$ & $0.491\pm0.086$ \\
204310 & $53.175481$ & $-27.792530$ & $3.7113$ & $3.34$ & $1.65_{-0.38}^{+0.54}$ & $215.7\pm83.5$ & $<0.175$ & $<0.171$ \\
204346 & $53.135140$ & $-27.792345$ & $3.1091$ & $3.16$ & $1.27_{-0.11}^{+0.15}$ & $429.8\pm42.0$ & $0.779\pm0.067$ & $<0.123$ \\
204449 & $53.160657$ & $-27.791837$ & $2.6348$ & $2.65$ & $1.75_{-0.01}^{+0.02}$ & $55.9\pm1.2$ & $3.112\pm0.213$ & $0.456\pm0.041$ \\
204702 & $53.136937$ & $-27.790756$ & $1.7473$ & $1.77$ & $1.70_{-0.07}^{+0.07}$ & $69.1\pm7.3$ & $0.957\pm0.317$ & $<0.191$ \\
205915 & $53.133262$ & $-27.786801$ & $2.9432$ & $2.96$ & $1.42_{-0.06}^{+0.06}$ & $285.6\pm34.1$ & $<1.009$ & $0.349\pm0.036$ \\
205932 & $53.117985$ & $-27.786608$ & $2.2375$ & $2.04$ & $1.44_{-0.03}^{+0.04}$ & $129.5\pm6.6$ & $2.597\pm0.033$ & $0.216\pm0.028$ \\
206579 & $53.181179$ & $-27.785152$ & $2.8115$ & $2.88$ & $1.60_{-0.10}^{+0.10}$ & $153.0\pm27.2$ & $<0.669$ & $0.087\pm0.041$ \\
207194 & $53.162801$ & $-27.782300$ & $2.3467$ & $2.41$ & $1.92_{-0.13}^{+0.17}$ & $34.2\pm7.2$ & $0.430\pm0.101$ & $<0.048$ \\
285081 & $53.164480$ & $-27.815769$ & $2.6646$ & $1.69$ & $1.56_{-0.19}^{+0.25}$ & $1139.5\pm132.4$ & $1.274\pm0.295$ & $<0.185$ \\
2022016 & $53.167470$ & $-27.795250$ & $1.7775$ & $0.00$ & $1.56_{-0.06}^{+0.06}$ & $88.4\pm24.7$ & $<2.551$ & $<0.207$ \\
2022040 & $53.166810$ & $-27.818560$ & $1.9267$ & $0.00$ & $1.63_{-0.02}^{+0.02}$ & $4.8\pm34.7$ & $<4.834$ & $0.248\pm0.039$ \\
2023007 & $53.182342$ & $-27.783406$ & $2.0709$ & $0.00$ & $1.33_{-0.06}^{+0.06}$ & $11.8\pm22.3$ & $<0.761$ & $<0.100$ \\
2023010 & $53.164408$ & $-27.838797$ & $3.5920$ & $0.00$ & $1.57_{-0.14}^{+0.16}$ & $121.8\pm8.7$ & $0.322\pm0.053$ & $<0.115$ \\
2023026 & $53.135280$ & $-27.814521$ & $2.8122$ & $0.00$ & $1.20_{-0.08}^{+0.09}$ & $123.5\pm9.9$ & $0.580\pm0.183$ & $0.254\pm0.038$ \\
\enddata
\tablecomments{Coordinates are in degrees. Equivalent widths are rest-frame values in \AA. Integrated line fluxes are in $10^{-18}\,{\rm erg\,s^{-1}\,cm^{-2}}$. Missing values are shown as \nodata.}
\end{deluxetable*}

\startlongtable
\begin{deluxetable*}{lccccccc}
\tabletypesize{\scriptsize}
\tablewidth{0pt}
\tablecaption{Derived galaxy properties.\label{tab:derived_properties}}
\tablehead{
\colhead{JADES ID} & \colhead{$\log_{10}(M_\star/M_\odot)$} & \colhead{MWA} & \colhead{$\tau_V$} & \colhead{$E(B-V)_\star$} & \colhead{${\rm SFR}_{\rm SED}$} & \colhead{${\rm SFR}_{\rm H\alpha}$} & \colhead{${\rm SFR}_{\rm Pa\gamma}$} \\
\colhead{---} & \colhead{---} & \colhead{Gyr} & \colhead{---} & \colhead{mag} & \colhead{$M_\odot\,{\rm yr}^{-1}$} & \colhead{$M_\odot\,{\rm yr}^{-1}$} & \colhead{$M_\odot\,{\rm yr}^{-1}$}
}
\startdata
083749 & $8.66_{-0.15}^{+0.14}$ & $0.68_{-0.34}^{+0.21}$ & $0.185_{-0.074}^{+0.232}$ & $0.050_{-0.020}^{+0.062}$ & $0.920_{-0.819}^{+1.416}$ & $0.966\pm0.038$ & $6.140\pm1.371$ \\
085339 & $8.88_{-0.21}^{+0.09}$ & $0.97_{-0.28}^{+0.10}$ & $0.129_{-0.052}^{+0.103}$ & $0.035_{-0.014}^{+0.028}$ & $2.222_{-0.923}^{+1.225}$ & $1.640\pm0.069$ & $2.902\pm0.437$ \\
085407 & $8.94_{-0.16}^{+0.14}$ & $0.93_{-0.34}^{+0.14}$ & $0.425_{-0.289}^{+0.539}$ & $0.114_{-0.077}^{+0.145}$ & $0.649_{-0.515}^{+3.349}$ & $<0.005$ & $<0.062$ \\
085997 & $8.44_{-0.12}^{+0.14}$ & $0.89_{-0.29}^{+0.14}$ & $0.159_{-0.110}^{+0.241}$ & $0.043_{-0.030}^{+0.065}$ & $0.213_{-0.146}^{+0.213}$ & $<0.018$ & $<0.228$ \\
086866 & $8.82_{-0.06}^{+0.19}$ & $0.60_{-0.22}^{+0.38}$ & $0.041_{-0.031}^{+0.505}$ & $0.011_{-0.008}^{+0.135}$ & $0.060_{-0.052}^{+0.172}$ & $<0.041$ & $<0.150$ \\
087622 & $8.48_{-0.25}^{+0.22}$ & $0.85_{-0.32}^{+0.15}$ & $0.084_{-0.052}^{+0.105}$ & $0.023_{-0.014}^{+0.028}$ & $0.708_{-0.404}^{+0.560}$ & $0.194\pm0.078$ & $<2.687$ \\
088115 & $8.28_{-0.13}^{+0.08}$ & $0.94_{-0.24}^{+0.12}$ & $0.229_{-0.124}^{+0.138}$ & $0.061_{-0.033}^{+0.037}$ & $0.051_{-0.043}^{+0.084}$ & $0.030\pm0.011$ & $0.214\pm0.062$ \\
089220 & $8.28_{-0.14}^{+0.13}$ & $0.96_{-0.25}^{+0.13}$ & $0.306_{-0.187}^{+0.216}$ & $0.082_{-0.050}^{+0.058}$ & $0.218_{-0.152}^{+0.132}$ & $<2.2e-06$ & $<0.270$ \\
089509 & $8.33_{-0.27}^{+0.27}$ & $0.77_{-0.35}^{+0.25}$ & $0.094_{-0.072}^{+0.203}$ & $0.025_{-0.019}^{+0.054}$ & $0.264_{-0.157}^{+0.497}$ & $0.149\pm0.015$ & $<0.989$ \\
090417 & $9.47_{-0.11}^{+0.07}$ & $0.82_{-0.47}^{+0.12}$ & $0.666_{-0.044}^{+0.035}$ & $0.179_{-0.012}^{+0.009}$ & $23.460_{-13.841}^{+3.543}$ & $6.182\pm0.417$ & $19.632\pm0.667$ \\
090735 & $9.08_{-0.12}^{+0.10}$ & $0.84_{-0.16}^{+0.15}$ & $0.085_{-0.032}^{+0.049}$ & $0.023_{-0.008}^{+0.013}$ & $3.368_{-0.664}^{+1.188}$ & $0.434\pm0.177$ & $3.904\pm0.400$ \\
092192 & $9.02_{-0.05}^{+0.06}$ & $0.91_{-0.16}^{+0.12}$ & $0.133_{-0.056}^{+0.106}$ & $0.036_{-0.015}^{+0.028}$ & $1.061_{-0.582}^{+0.304}$ & $0.037\pm0.002$ & $0.775\pm0.171$ \\
093268 & $8.57_{-0.16}^{+0.10}$ & $0.91_{-0.25}^{+0.12}$ & $0.113_{-0.075}^{+0.168}$ & $0.030_{-0.020}^{+0.045}$ & $0.147_{-0.090}^{+0.181}$ & $0.045\pm0.018$ & $<0.301$ \\
094972 & $8.73_{-0.30}^{+0.10}$ & $0.81_{-0.34}^{+0.15}$ & $0.115_{-0.066}^{+0.161}$ & $0.031_{-0.018}^{+0.043}$ & $0.792_{-0.369}^{+0.533}$ & \nodata & $<1.634$ \\
097581 & $8.94_{-0.44}^{+0.09}$ & $0.56_{-0.39}^{+0.29}$ & $0.266_{-0.209}^{+0.036}$ & $0.071_{-0.056}^{+0.010}$ & $2.735_{-1.328}^{+0.448}$ & $1.942\pm0.073$ & \nodata \\
097858 & $8.39_{-0.22}^{+0.23}$ & $0.69_{-0.32}^{+0.25}$ & $0.112_{-0.042}^{+0.103}$ & $0.030_{-0.011}^{+0.028}$ & $1.129_{-0.384}^{+0.645}$ & $<6.1e-07$ & $1.730\pm0.422$ \\
099110 & $8.71_{-0.15}^{+0.12}$ & $0.90_{-0.27}^{+0.15}$ & $0.103_{-0.051}^{+0.076}$ & $0.028_{-0.014}^{+0.020}$ & $0.975_{-0.335}^{+0.511}$ & $<0.093$ & $<1.903$ \\
115313 & $7.73_{-0.61}^{+0.33}$ & $0.70_{-0.48}^{+0.25}$ & $0.057_{-0.021}^{+0.041}$ & $0.015_{-0.006}^{+0.011}$ & $0.673_{-0.315}^{+0.230}$ & $0.567\pm0.052$ & \nodata \\
115803 & $8.67_{-0.13}^{+0.13}$ & $0.77_{-0.22}^{+0.18}$ & $0.042_{-0.029}^{+0.042}$ & $0.011_{-0.008}^{+0.011}$ & $0.137_{-0.083}^{+0.358}$ & $0.015\pm0.007$ & $<0.332$ \\
116235 & $7.95_{-0.40}^{+0.18}$ & $0.71_{-0.42}^{+0.17}$ & $0.034_{-0.019}^{+0.023}$ & $0.009_{-0.005}^{+0.006}$ & $0.327_{-0.103}^{+0.171}$ & $0.257\pm0.024$ & $0.642\pm0.272$ \\
116687 & $8.97_{-0.26}^{+0.11}$ & $0.76_{-0.37}^{+0.22}$ & $0.267_{-0.231}^{+0.164}$ & $0.072_{-0.062}^{+0.044}$ & $0.065_{-0.059}^{+0.405}$ & $0.084\pm0.012$ & $0.518\pm0.244$ \\
121632 & $8.07_{-0.27}^{+0.09}$ & $0.84_{-0.13}^{+0.09}$ & $0.160_{-0.057}^{+0.078}$ & $0.043_{-0.015}^{+0.021}$ & $1.821_{-1.101}^{+0.723}$ & $<0.549$ & $1.832\pm0.352$ \\
123725 & $8.03_{-0.14}^{+0.11}$ & $0.98_{-0.20}^{+0.11}$ & $0.403_{-0.209}^{+0.145}$ & $0.108_{-0.056}^{+0.039}$ & $0.079_{-0.053}^{+0.037}$ & $<0.007$ & $<0.081$ \\
123903 & $8.35_{-0.52}^{+0.21}$ & $0.86_{-0.41}^{+0.18}$ & $0.223_{-0.079}^{+0.204}$ & $0.060_{-0.021}^{+0.055}$ & $1.036_{-0.395}^{+0.870}$ & $0.567\pm0.058$ & $<2.169$ \\
124533 & $8.61_{-0.18}^{+0.14}$ & $0.87_{-0.29}^{+0.16}$ & $0.326_{-0.184}^{+0.252}$ & $0.087_{-0.049}^{+0.068}$ & $1.304_{-0.705}^{+0.746}$ & $0.179\pm0.085$ & $1.450\pm0.213$ \\
125100 & $8.64_{-0.16}^{+0.09}$ & $0.76_{-0.26}^{+0.18}$ & $0.391_{-0.198}^{+0.185}$ & $0.105_{-0.053}^{+0.050}$ & $0.110_{-0.099}^{+0.388}$ & $<0.033$ & $<0.411$ \\
125748 & $9.63_{-0.11}^{+0.14}$ & $0.92_{-0.23}^{+0.13}$ & $0.330_{-0.216}^{+0.421}$ & $0.089_{-0.058}^{+0.113}$ & $1.526_{-1.251}^{+5.601}$ & $<0.163$ & $2.786\pm0.698$ \\
126238 & $7.97_{-0.48}^{+0.26}$ & $0.63_{-0.39}^{+0.25}$ & $0.112_{-0.058}^{+0.091}$ & $0.030_{-0.016}^{+0.024}$ & $1.167_{-0.670}^{+0.600}$ & $0.814\pm0.079$ & $5.003\pm2.378$ \\
191229 & $8.23_{-0.18}^{+0.13}$ & $0.80_{-0.23}^{+0.20}$ & $0.196_{-0.149}^{+0.121}$ & $0.053_{-0.040}^{+0.033}$ & $0.145_{-0.082}^{+0.088}$ & $0.042\pm0.006$ & $<0.182$ \\
191896 & $8.40_{-0.38}^{+0.20}$ & $0.77_{-0.37}^{+0.22}$ & $0.177_{-0.071}^{+0.117}$ & $0.047_{-0.019}^{+0.032}$ & $0.554_{-0.310}^{+0.574}$ & $<0.028$ & $<0.149$ \\
192046 & $8.84_{-0.15}^{+0.11}$ & $0.87_{-0.33}^{+0.15}$ & $0.226_{-0.136}^{+0.339}$ & $0.061_{-0.036}^{+0.091}$ & $1.857_{-1.056}^{+0.915}$ & $0.076\pm0.030$ & $<0.365$ \\
192594 & $9.21_{-0.16}^{+0.10}$ & $1.02_{-0.20}^{+0.10}$ & $0.727_{-0.139}^{+0.145}$ & $0.195_{-0.037}^{+0.039}$ & $0.920_{-0.522}^{+0.379}$ & $<0.445$ & $<0.661$ \\
193779 & $9.20_{-0.05}^{+0.17}$ & $0.72_{-0.14}^{+0.20}$ & $0.141_{-0.080}^{+0.015}$ & $0.038_{-0.021}^{+0.004}$ & $10.914_{-4.628}^{+2.007}$ & $<1.654$ & $3.457\pm0.571$ \\
194960 & $9.67_{-0.06}^{+0.07}$ & $0.83_{-0.20}^{+0.16}$ & $0.211_{-0.088}^{+0.132}$ & $0.057_{-0.024}^{+0.035}$ & $3.154_{-1.230}^{+2.484}$ & $0.735\pm0.160$ & $6.030\pm0.684$ \\
195115 & $8.37_{-0.15}^{+0.07}$ & $0.75_{-0.29}^{+0.17}$ & $0.200_{-0.017}^{+0.048}$ & $0.054_{-0.005}^{+0.013}$ & $8.838_{-5.750}^{+1.313}$ & $<2.129$ & $9.365\pm1.906$ \\
195924 & $8.85_{-0.16}^{+0.14}$ & $0.87_{-0.20}^{+0.13}$ & $0.121_{-0.074}^{+0.103}$ & $0.032_{-0.020}^{+0.028}$ & $1.055_{-0.455}^{+0.905}$ & $<0.246$ & $0.650\pm0.228$ \\
196204 & $7.84_{-0.51}^{+0.31}$ & $0.43_{-0.35}^{+0.46}$ & $0.325_{-0.139}^{+0.079}$ & $0.087_{-0.037}^{+0.021}$ & $1.571_{-1.064}^{+1.542}$ & $0.246\pm0.090$ & $1.010\pm0.252$ \\
197390 & $8.92_{-0.09}^{+0.05}$ & $0.99_{-0.25}^{+0.13}$ & $0.094_{-0.062}^{+0.104}$ & $0.025_{-0.017}^{+0.028}$ & $0.310_{-0.202}^{+0.366}$ & $0.174\pm0.021$ & $0.642\pm0.252$ \\
198754 & $9.21_{-0.03}^{+0.03}$ & $0.91_{-0.14}^{+0.08}$ & $0.261_{-0.062}^{+0.057}$ & $0.070_{-0.017}^{+0.015}$ & $0.526_{-0.212}^{+0.765}$ & $<0.655$ & $<0.173$ \\
201169 & $8.83_{-0.04}^{+0.03}$ & $0.70_{-0.10}^{+0.09}$ & $0.347_{-0.042}^{+0.051}$ & $0.093_{-0.011}^{+0.014}$ & $1.934_{-0.860}^{+0.426}$ & $0.133\pm0.049$ & $2.955\pm0.211$ \\
202445 & $8.86_{-0.47}^{+0.29}$ & $0.67_{-0.34}^{+0.28}$ & $0.154_{-0.080}^{+0.082}$ & $0.041_{-0.021}^{+0.022}$ & $1.973_{-1.036}^{+1.112}$ & $0.541\pm0.087$ & $<1.745$ \\
202882 & $9.02_{-0.09}^{+0.08}$ & $0.88_{-0.23}^{+0.15}$ & $0.026_{-0.020}^{+0.054}$ & $0.007_{-0.005}^{+0.015}$ & $0.255_{-0.155}^{+0.515}$ & $<0.253$ & $0.950\pm0.157$ \\
203991 & $9.19_{-0.05}^{+0.04}$ & $0.54_{-0.19}^{+0.35}$ & $0.312_{-0.036}^{+0.043}$ & $0.084_{-0.010}^{+0.011}$ & $1.343_{-0.404}^{+0.445}$ & $0.091\pm0.021$ & $1.005\pm0.176$ \\
204310 & $8.11_{-0.21}^{+0.13}$ & $0.93_{-0.28}^{+0.13}$ & $0.086_{-0.055}^{+0.100}$ & $0.023_{-0.015}^{+0.027}$ & $0.277_{-0.158}^{+0.254}$ & $<0.139$ & $<4.002$ \\
204346 & $8.50_{-0.06}^{+0.14}$ & $0.48_{-0.18}^{+0.48}$ & $0.269_{-0.137}^{+0.038}$ & $0.072_{-0.037}^{+0.010}$ & $1.160_{-0.553}^{+0.401}$ & $0.471\pm0.041$ & $<1.886$ \\
204449 & $9.84_{-0.02}^{+0.02}$ & $0.33_{-0.10}^{+0.67}$ & $0.352_{-0.027}^{+0.035}$ & $0.094_{-0.007}^{+0.009}$ & $11.398_{-3.381}^{+4.882}$ & $1.353\pm0.093$ & $4.694\pm0.427$ \\
204702 & $9.26_{-0.07}^{+0.05}$ & $0.81_{-0.11}^{+0.11}$ & $0.372_{-0.048}^{+0.034}$ & $0.100_{-0.013}^{+0.009}$ & $2.043_{-0.847}^{+0.566}$ & $0.155\pm0.051$ & $<0.723$ \\
205915 & $8.63_{-0.14}^{+0.22}$ & $0.68_{-0.43}^{+0.24}$ & $0.246_{-0.127}^{+0.072}$ & $0.066_{-0.034}^{+0.019}$ & $1.369_{-0.603}^{+0.513}$ & $<0.525$ & $4.691\pm0.479$ \\
205932 & $8.61_{-0.07}^{+0.06}$ & $0.37_{-0.12}^{+0.46}$ & $0.072_{-0.004}^{+0.006}$ & $0.019_{-0.001}^{+0.002}$ & $11.118_{-5.031}^{+0.908}$ & $0.604\pm0.008$ & $1.501\pm0.195$ \\
206579 & $9.15_{-0.07}^{+0.08}$ & $0.65_{-0.24}^{+0.31}$ & $0.501_{-0.083}^{+0.103}$ & $0.134_{-0.022}^{+0.028}$ & $1.615_{-1.040}^{+0.834}$ & $<0.384$ & $1.052\pm0.495$ \\
207194 & $8.90_{-0.23}^{+0.16}$ & $0.85_{-0.37}^{+0.15}$ & $0.311_{-0.185}^{+0.169}$ & $0.083_{-0.050}^{+0.045}$ & $0.107_{-0.094}^{+0.346}$ & $0.137\pm0.032$ & $<0.371$ \\
285081 & $8.55_{-0.18}^{+0.05}$ & $0.92_{-0.26}^{+0.10}$ & $0.173_{-0.033}^{+0.145}$ & $0.046_{-0.009}^{+0.039}$ & $2.743_{-2.137}^{+0.722}$ & $0.491\pm0.114$ & $<1.956$ \\
2022016 & $9.80_{-0.01}^{+0.01}$ & $0.86_{-0.10}^{+0.14}$ & $0.459_{-0.023}^{+0.025}$ & $0.123_{-0.006}^{+0.007}$ & $4.165_{-0.625}^{+1.934}$ & $<0.464$ & $<0.817$ \\
2022040 & $9.69_{-0.03}^{+0.03}$ & $0.83_{-0.09}^{+0.15}$ & $0.271_{-0.030}^{+0.036}$ & $0.073_{-0.008}^{+0.010}$ & $5.519_{-3.042}^{+0.966}$ & $<0.919$ & $1.196\pm0.189$ \\
2023007 & $9.23_{-0.03}^{+0.04}$ & $0.86_{-0.15}^{+0.06}$ & $0.468_{-0.044}^{+0.048}$ & $0.126_{-0.012}^{+0.013}$ & $1.731_{-0.806}^{+0.746}$ & $<0.203$ & $<0.577$ \\
2023010 & $8.59_{-0.66}^{+0.10}$ & $0.83_{-0.29}^{+0.14}$ & $0.066_{-0.039}^{+0.068}$ & $0.018_{-0.010}^{+0.018}$ & $0.197_{-0.148}^{+0.329}$ & $0.232\pm0.039$ & $<2.492$ \\
2023026 & $9.86_{-0.02}^{+0.02}$ & $0.99_{-0.34}^{+0.02}$ & $0.419_{-0.016}^{+0.019}$ & $0.112_{-0.004}^{+0.005}$ & $9.132_{-2.962}^{+15.104}$ & $0.312\pm0.098$ & $3.068\pm0.460$ \\
\enddata
\tablecomments{MWA is the mass-weighted stellar age in Gyr. $\tau_V$ is the dimensionless SED parameter \texttt{dust2}; $E(B-V)_\star$ is in magnitudes. All SFRs are in $M_\odot\,{\rm yr^{-1}}$. The SED SFR uses the first SFH bin, corresponding to 0--5 Myr under the adopted binning. H$\alpha$ SFRs include the adopted correction $\exp(0.82\tau_V)$; Pa$\gamma$ SFRs are uncorrected. H$\alpha$ errors and limits include line-flux uncertainty only. For poor line detections ($F/\sigma_F<2$, including negative fitted fluxes), $<$ denotes a noise-based $2\sigma$ upper limit calculated from $2\sigma_F$. Missing or invalid flux uncertainties do not yield upper limits.}
\end{deluxetable*}

\subsection{Equivalent Width versus Break Strength}

\begin{figure}[!ht]
    \centering
    \includegraphics[width=\columnwidth]{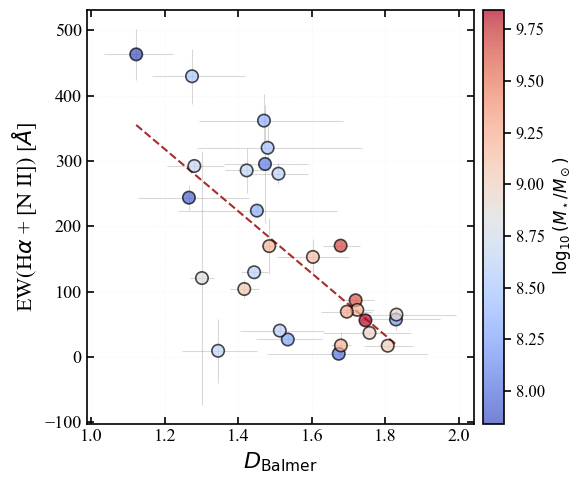}
    \caption{
        Rest-frame equivalent width of H$\alpha$+[N\,\textsc{ii}]
        as a function of the $D_{\rm Balmer}$ index for
        29 galaxies. Points are colored by stellar mass,
        and the dashed line shows a linear fit illustrating
        the strong anti-correlation ($r=-0.680$,
        $p=4.91\times10^{-5}$). Horizontal error bars show
        the 16th--84th percentile intervals for
        $D_{\rm Balmer}$, while vertical error bars show
        the equivalent-width uncertainties obtained by
        combining the component uncertainties in quadrature.
        This trend indicates decreasing recent star formation
        activity with increasing Balmer-break strength.
    }
    \label{fig:ew_DBalmer}
\end{figure}

To explore the connection between recent star formation
activity and stellar population age, we compared the
rest-frame equivalent width of the blended
H$\alpha$+[N\,\textsc{ii}] line with the previously calculated
$D_{\rm Balmer}$ break strength. We retained the original,
visually inspected sample for this comparison, with one
entry per galaxy and updated break measurements and
uncertainties. No additional cut on the absolute
$D_{\rm Balmer}$ uncertainty was imposed. Our final sample
contains 29 galaxies with measurements of both diagnostics,
as well as stellar masses derived from photometric SED
fitting with {\tt PROSPECTOR}.

The resulting trend is a clear negative correlation between
$D_{\rm Balmer}$ and the rest-frame equivalent width of the
blended H$\alpha$+[N\,\textsc{ii}] line, as seen in
Figure~\ref{fig:ew_DBalmer}. We find that the trend is
statistically significant, with a Pearson correlation
coefficient of $r=-0.680$ and a $p$-value of
$4.91\times10^{-5}$. The Spearman correlation coefficient
is $\rho=-0.616$, with $p=3.77\times10^{-4}$.
This trend is consistent with the expected relationship
between stellar population age and hydrogen recombination-line
strength. Galaxies with younger stellar populations are
more likely to be actively forming stars, resulting in
higher equivalent widths and smaller $D_{\rm Balmer}$ values,
while galaxies with older stellar populations tend to have
stronger breaks and smaller equivalent widths. Stellar
masses, shown by the color coding, span approximately two
orders of magnitude.

We also note the absence of galaxies with strong Balmer breaks and high H$\alpha$+[N\,II] equivalent widths in our sample. However, the absence of these galaxies does not rule out the possibility of recurring star formation. A renewed period of star formation could increase the nebular emission but also weaken the Balmer break, moving the galaxy diagonally instead of vertically in this plot. Our current data does not show strong evidence of mini-quenching. 

\subsection{Break Strength vs.\ Stellar Population Age}

To understand the relationship between the Balmer-break strength, $D_{\rm Balmer}$, and the stellar population age, we studied the correlation between $D_{\rm Balmer}$ and the mass-weighted stellar age. This relationship was computed for galaxies with both star formation histories and spectral coverage. The mass-weighted age (MWA) was derived by integrating the age of each time bin, weighted by the stellar mass formed in that bin:
\begin{equation}
    \mathrm{MWA} = \frac{\sum_i \mathrm{SFR}_i \cdot \Delta t_i \cdot t_i}{\sum_i \mathrm{SFR}_i \cdot \Delta t_i},
\end{equation}
where $\mathrm{SFR}_i$ is the star formation rate in bin $i$, $\Delta t_i$ is the duration of the bin, and $t_i$ is the lookback time corresponding to the bin center. This yields a final sample of 57 galaxies.

\begin{figure}[!ht]
    \centering
    \includegraphics[width=0.95\columnwidth]{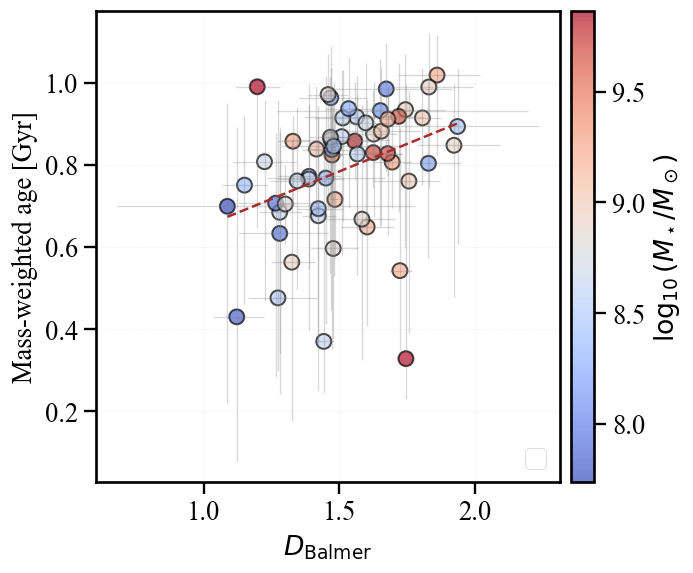}
    \caption{Mass-weighted stellar age versus $D_{\rm Balmer}$. Points are color-coded by stellar mass, and the dashed line shows the linear regression. We find a Spearman correlation of $\rho=0.412$ ($p=1.44\times10^{-3}$) and a Pearson correlation of $r=0.351$ ($p=7.44\times10^{-3}$). Error bars show the propagated measurement uncertainties.}
    \label{fig:mass_weighted_age_vs_d4000}
\end{figure}

As shown in Figure \ref{fig:mass_weighted_age_vs_d4000}, $D_{\rm Balmer}$ shows a positive trend with the mass-weighted age. In other words, galaxies with larger $D_{\rm Balmer}$ values tend to host older stellar populations. This trend is consistent with stellar population synthesis models, which shows that the Balmer absorption features strengthen at ages of $\sim 0.1$--$1$ Gyr and that the Balmer break at 3646~\AA{} evolves toward the classical 4000~\AA{} break as the stellar population ages \citep{bruzual_stellar_2003}. In low-redshift SDSS studies, the 4000~\AA{} break strength has been used to differentiate galaxies dominated by old stellar populations from those that have experienced a recent burst of star formation and to constrain mean stellar population ages \citep{kauffmann_stellar_2003}. The observed correlation therefore supports the use of $D_{\rm Balmer}$ as an empirical tracer of stellar population age at cosmic noon.

Several galaxies lie noticeably below the overall
$D_{\rm Balmer}$--MWA relation. The most prominent cases are
JADES IDs 204449, 205932, and 203991, which have relatively
large break strengths ($D_{\rm Balmer}=1.75$, 1.44, and 1.72,
respectively) but low mass-weighted ages of approximately
0.33, 0.37, and 0.54 Gyr. Their Prospector star-formation
histories show substantial recent star formation followed by
declining activity. Because the mass-weighted age is weighted
by the stellar mass formed in each time bin, recent episodes
can keep the inferred age relatively young even when the
continuum already shows a pronounced Balmer break. The
uncertainties on these ages are also broad, particularly for
204449 and 205932.

Conversely, JADES ID 2023026 has a relatively weak break
($D_{\rm Balmer}=1.20$) but a larger mass-weighted age of
approximately 0.99 Gyr. Its star-formation history contains
an older underlying population while retaining substantial
recent star formation, which can weaken the observed break.
These cases illustrate that $D_{\rm Balmer}$ and mass-weighted
age trace related but different aspects of the star-formation
history. The deviations do not invalidate the
overall positive trend, but show that the break strength is
also sensitive to the timing of recent star formation.

\subsection{SED and Emission Line Based Star Formation Rate Comparisons}

\begin{figure*}[!ht]
\centering
\includegraphics[width=1\textwidth]{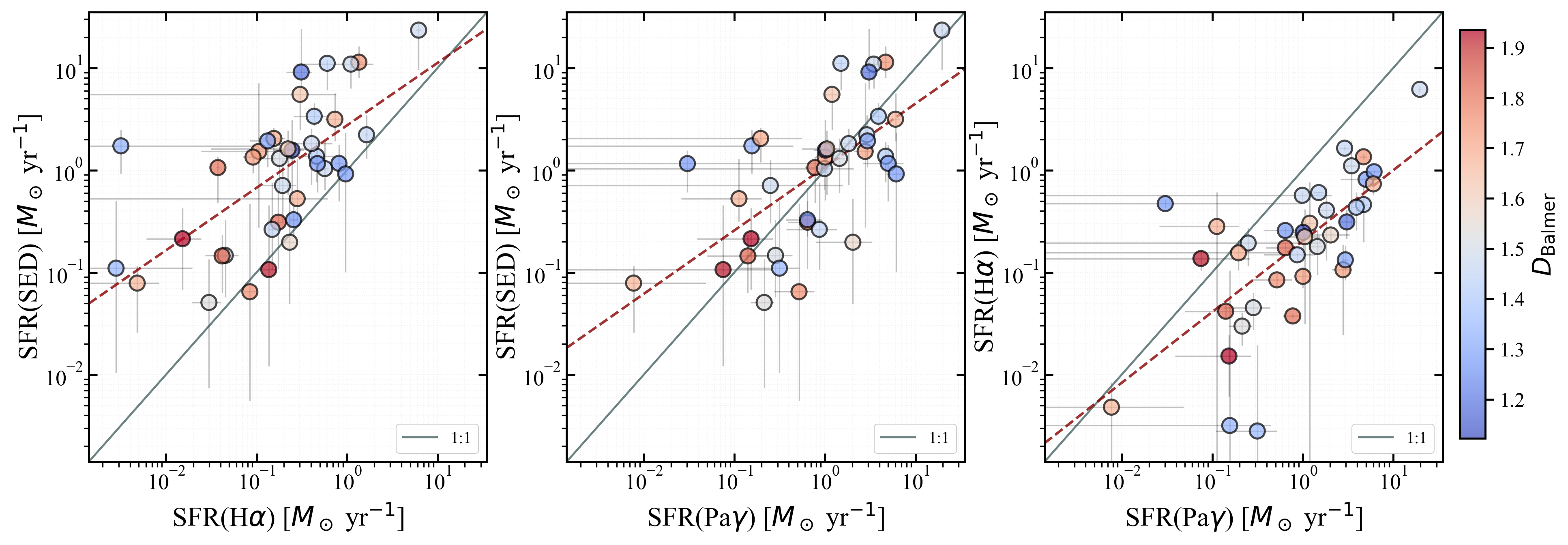}
\caption{
Comparison of star formation rates derived from SED fitting and
emission lines for 38 unique galaxies. (a) SFR(SED) versus
SFR(H$\alpha$), (b) SFR(SED) versus SFR(Pa$\gamma$), and
(c) SFR(H$\alpha$) versus SFR(Pa$\gamma$). Points are color-coded
by $D_{\rm Balmer}$, dashed lines indicate best-fit relations,
and solid lines show the one-to-one relation. Error bars show
the propagated measurement uncertainties.
}
\label{fig:sfr_comparisons}
\end{figure*}

Next, we assessed the consistency between the different star
formation rate (SFR) diagnostics derived from {\tt GELATO} and
{\tt PROSPECTOR}. Figure~\ref{fig:sfr_comparisons} presents three
comparisons: SFR$_{\rm SED}$ versus SFR(H$\alpha$),
SFR$_{\rm SED}$ versus SFR(Pa$\gamma$), and SFR(H$\alpha$) versus
SFR(Pa$\gamma$). All SFRs are shown on a logarithmic scale and
are color-coded by the $D_{\rm Balmer}$ index to show trends with
stellar population age.

The SFR determined from SED fitting, SFR$_{\rm SED}$, is given by
the SFR in the first {\tt PROSPECTOR} SFH bin, corresponding to
the 0--5~Myr interval under the adopted binning. This rate
therefore traces very recent star formation. The emission-line
SFRs also trace short timescales; for example, H$\alpha$ is
sensitive to star formation over approximately 10~Myr. These
different timescales provide a direct comparison between the
SED-based and emission-line indicators while retaining their
distinct physical sensitivities.

We adopted the H$\alpha$ calibration from
\citet{kennicutt_star_1998}, scaled to a
\citet{chabrier_galactic_2003} IMF following
\citet{Belfiore2023}:
\begin{equation}
\mathrm{SFR(H\alpha)}
=10^{-41.26} L_{\mathrm{H}\alpha}
\quad [\mathrm{M}_{\odot}\,\mathrm{yr}^{-1}],
\end{equation}
where $L_{\mathrm{H}\alpha}$ is in units of
erg~s$^{-1}$.

For Pa$\gamma$, we rescaled this calibration using the theoretical
Case B recombination ratio
$L_{\mathrm{H}\alpha}/L_{\mathrm{Pa}\gamma}=31.6$ for
$T_e=10^4$~K and $n_e=100$~cm$^{-3}$
\citep{osterbrock_astrophysics_1989}. The resulting calibration is
\begin{equation}
\mathrm{SFR(Pa\gamma)}
=\frac{10^{-41.26}}{0.0316}L_{\mathrm{Pa}\gamma},
\end{equation}
where $L_{\mathrm{Pa}\gamma}$ is in units of erg~s$^{-1}$.

The luminosities were calculated using luminosity distances from
our adopted cosmology. For H$\alpha$, we applied an attenuation
correction using the Prospector diffuse-dust parameter,
\begin{equation}
L_{\mathrm{H}\alpha,\mathrm{corr}}
=
L_{\mathrm{H}\alpha,\mathrm{obs}}
\exp(0.82\,\mathrm{dust2}).
\end{equation}
The corrected H$\alpha$ luminosity was then used to calculate
SFR(H$\alpha$). The Pa$\gamma$ SFRs were left uncorrected for
dust attenuation.

The SFRs derived from {\tt PROSPECTOR} SED fitting, H$\alpha$,
and Pa$\gamma$ emission lines show positive correlations. This
indicates broad consistency among the three SFR indicators. The
SED-based and Pa$\gamma$ SFRs are systematically larger than the
dust-corrected H$\alpha$ SFR. The scatter, especially at lower SFR,
likely reflects a combination of dust attenuation uncertainties,
line-measurement uncertainties, aperture differences, and the
different timescales traced by each diagnostic. Overall, the
agreement supports the use of emission-line SFRs as a consistency
check on recent star formation activity in the sample.

\subsection{Dust Attenuation from SED Fitting}

\begin{figure}[!ht]
\centering
\includegraphics[width=\columnwidth]{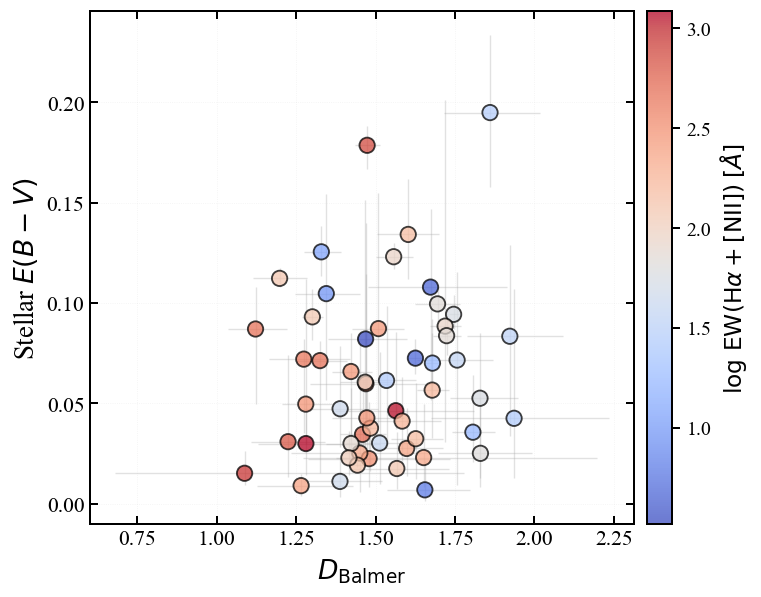}
\caption{Stellar $E(B-V)$ derived from {\tt PROSPECTOR} SED fitting versus Balmer-break strength ($D_{\rm Balmer}$). Color indicates the equivalent width of H$\alpha$+[N\,\textsc{ii}]. Error bars show the propagated measurement uncertainties.}
\label{fig:e(b-v)}
\end{figure}

To further characterize dust attenuation for our sample, we calculated the stellar color excess $E(B-V)_{\rm star}$ from {\tt PROSPECTOR} using the relation
\begin{equation}
    E(B-V)_{\rm star} = \frac{1.086\,\tau_V}{R'_V},    
\end{equation}
adopting $R'_V=4.05$, consistent with the \citet{calzetti_dust_2000} attenuation law. Here, $\tau_V$ is the effective dust-attenuation optical depth in the $V$ band inferred from the SED fit.

For our sample, 54 galaxies yield reliable measurements of both $E(B-V)_{\rm star}$ and $D_{\rm Balmer}$. The distribution has a median of $E(B-V)_{\rm star}=0.0546$ mag and a mean of $0.0621$ mag. As shown in Figure~\ref{fig:e(b-v)}, we find no statistically significant correlation between $D_{\rm Balmer}$ and $E(B-V)_{\rm star}$, with a Spearman correlation coefficient of $\rho=0.148$ and a $p$-value of $0.284$.

The large scatter and weak correlation support the conclusion that dust attenuation does not strongly drive the $D_{\rm Balmer}$ trends found in our sample. This suggests that the stronger correlations between $D_{\rm Balmer}$ and other galaxy properties, such as the equivalent-width trend shown in Figure~\ref{fig:ew_DBalmer}, are unlikely to be caused primarily by dust attenuation. Instead, $D_{\rm Balmer}$ remains a useful empirical tracer of stellar population age and recent star-formation history in this sample.

\subsection{Star Formation Rate Offset from the Main Sequence versus Balmer Break Strength}

\begin{figure*}[!ht]
\centering
\includegraphics[width=1\linewidth]{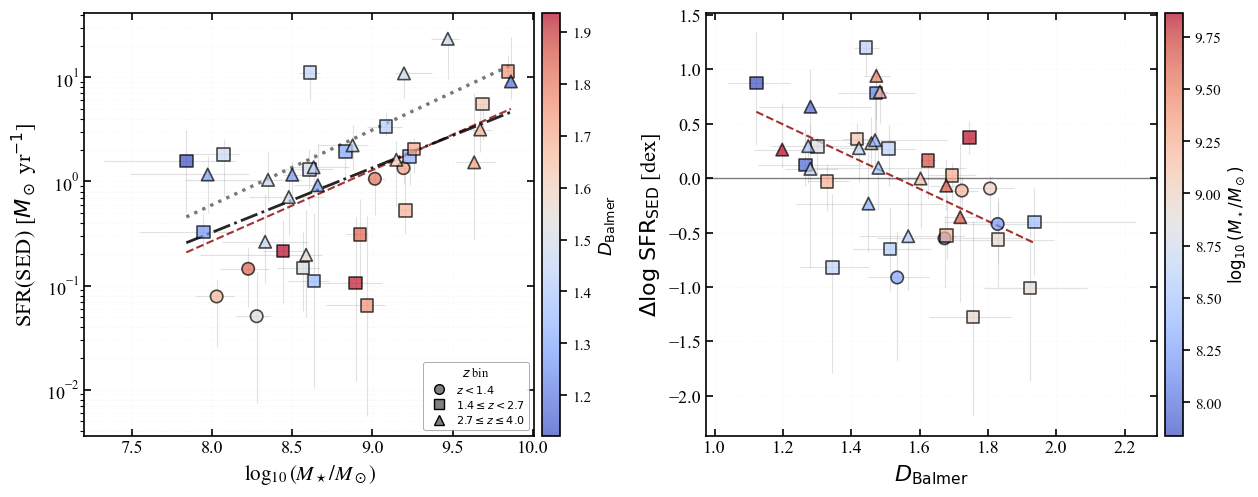}
\caption{\textbf{Left:} Star formation rate from SED fitting, $\mathrm{SFR}_{\rm SED}$, as a function of stellar mass. Points are colored by Balmer-break strength, $D_{\rm Balmer}$, and marker shapes indicate redshift bin: circles for $z<1.4$, squares for $1.4\leq z<2.7$, and triangles for $2.7\leq z\leq4.0$. The dashed dark-red line shows the best-fit main sequence of this sample. The black dash-dotted and gray dotted lines show the redshift-dependent star-forming main-sequence relations from \citet{clarke2024sfmsjadesceers} for $1.4\leq z<2.7$ and $2.7\leq z\leq4.0$, respectively. \textbf{Right:} Offset from the best-fit main sequence, $\Delta\log \mathrm{SFR}_{\rm SED}$, versus Balmer-break strength, $D_{\rm Balmer}$. Color indicates stellar mass, $\log(M_\star/M_\odot)$. The dashed line shows a linear fit; a negative trend is observed. Error bars show the propagated measurement uncertainties.}
\label{fig:residuals_D4000}
\end{figure*}

Figure~\ref{fig:residuals_D4000} shows the relationship between star formation, stellar mass, and $D_{\rm Balmer}$ using the SED-based star formation rates. In the left panel, we define the star-forming main sequence for our sample by fitting $\log(\mathrm{SFR}_{\rm SED})$ as a function of the stellar mass, $\log M_\star$. We also show the best-fit relations from \citet{clarke2024sfmsjadesceers} for the relevant redshift bins to compare our sample with published JWST main-sequence measurements. The right panel then uses our best-fit star-forming main sequence to calculate the offset, $\Delta\log \mathrm{SFR}_{\rm SED}$, as a function of $D_{\rm Balmer}$. Positive residuals correspond to galaxies above the main sequence, while negative residuals indicate that a galaxy has a lower star formation rate than expected for its stellar mass. The sample includes 38 galaxies with valid measurements of $\mathrm{SFR}_{\rm SED}$, stellar mass, and $D_{\rm Balmer}$.

The left panel shows that $\mathrm{SFR}_{\rm SED}$ increases with stellar mass, with a Pearson correlation coefficient of $r=0.548$, a $p$-value of $3.67\times10^{-4}$, and a best-fit slope of $0.681$. The right panel shows a clear negative correlation between $D_{\rm Balmer}$ and the offset from the main sequence, with a Spearman correlation coefficient of $\rho=-0.555$ and a $p$-value of $2.96\times10^{-4}$. This suggests that galaxies with large Balmer breaks lie further below the main sequence, while galaxies with smaller break strengths lie closer to the main sequence.

\begin{figure*}[!ht]
\centering
\includegraphics[width=1\linewidth]{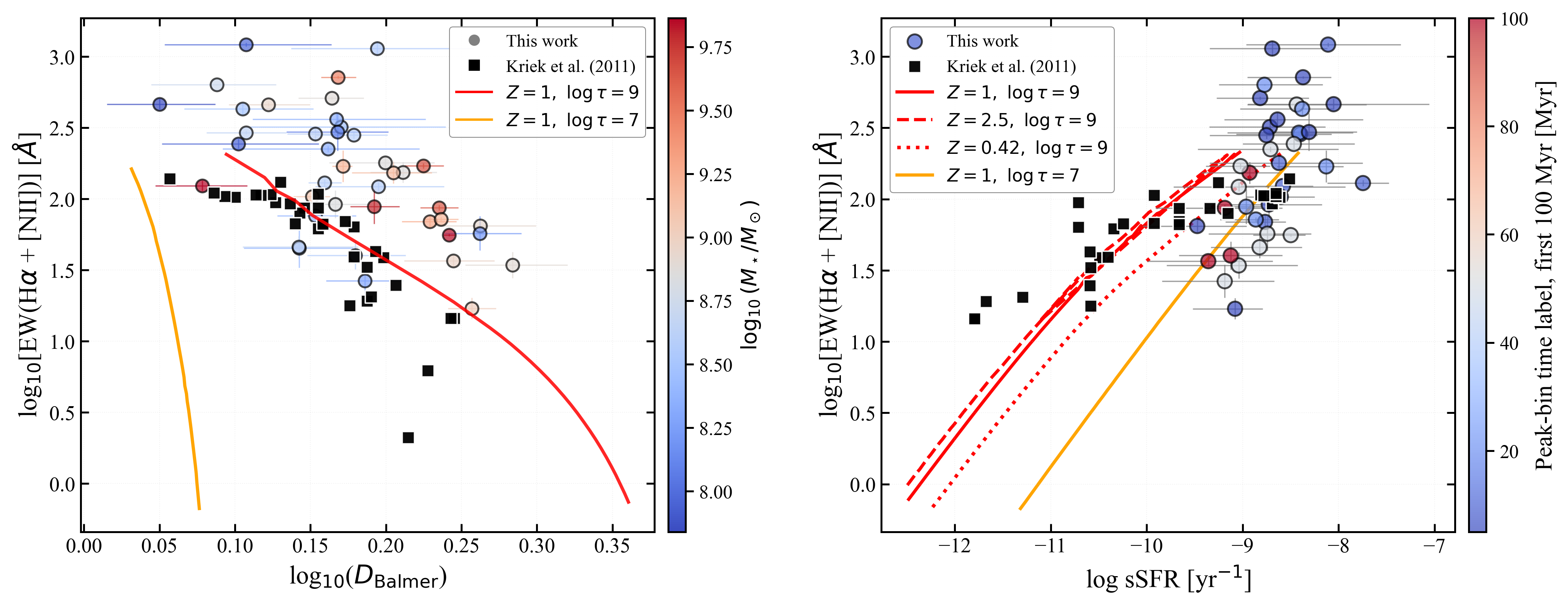}
\caption{\textbf{Left:} Rest-frame equivalent width of H$\alpha$+[N\,\textsc{ii}] versus the $D_{\rm Balmer}$ index for our sample of $N=41$ galaxies. Points are color-coded by stellar mass, with error bars showing the uncertainties in both quantities. Filled black squares show the $z<2$ comparison sample from \citet{kriek_h_2011}. The solid red line shows a synthetic SFH model with $Z=1$ and $\log\tau=9$, illustrating the expected evolution from high equivalent width and weak break strength toward low equivalent width and strong break strength. The orange curve shows a more rapidly declining model with $\log\tau=7$. Both quantities are plotted logarithmically. \textbf{Right:} H$\alpha$+[N\,\textsc{ii}] equivalent width versus SED-derived specific star formation rate (sSFR) for 38 galaxies. Points are color-coded by the peak-bin time label among the first five SFH bins, corresponding to the first 100 Myr. Horizontal error bars show the propagated sSFR uncertainties, while vertical error bars show the equivalent-width uncertainties. Solid, dashed, and dotted red lines show SFH model predictions with $Z=1$, $Z=2.5$, and $Z=0.42$, respectively, all with $\log\tau=9$. The orange curve shows a more rapidly declining model with $\log\tau=7$. Our galaxies follow a similar overall EW--break trend to the comparison sample but generally show larger equivalent widths at a given break strength and extend to higher sSFRs.}
\label{fig:d4000_rew}
\end{figure*}

\section{Discussion} \label{sec:discussion}

\subsection{Quenching Timescales and Recent Star Formation Suppression}

One of the most significant results from our analysis is the strong anti-correlation observed between H$\alpha$+[N\,\textsc{ii}] equivalent width and the $D_{\rm Balmer}$ index in galaxies at cosmic noon. This indicates that galaxies with stronger Balmer breaks have weaker nebular emission and lower recent star formation activity. Effectively, these galaxies span a range from the star-forming regime, marked by small $D_{\rm Balmer}$ and large H$\alpha$+[N\,\textsc{ii}] EW, to a more suppressed regime, marked by large $D_{\rm Balmer}$ and small H$\alpha$+[N\,\textsc{ii}] EW. This behavior is consistent with recent or rapid suppression of star formation, since H$\alpha$ emission fades on timescales of a few to tens of Myr after massive star formation declines, while the Balmer break strengthens as intermediate-age stars begin to dominate the continuum \citep{sun_evolution_2024}. Such a trajectory is expected for galaxies moving away from the blue star-forming population toward the red, quiescent population. Possible physical drivers include gas depletion, AGN feedback, starburst-driven outflows, and environmental gas removal \citep{carnall_inferring_2018,singh_quenching_2025}. Our data support this transition picture, while the specific quenching mechanism likely varies across the sample.

This interpretation is supported by the inclusion of data from \citet{kriek_h_2011}, which show a similar trend at lower redshift ($z < 2.0$), and by synthetic star formation history tracks that follow the same general trajectory in the $\log_{10}(D_{\rm Balmer})$ versus $\log_{10}({\rm EW[H}\alpha + {\rm [N\,\textsc{ii}]})$ plane. As shown in the left panel of Fig.~\ref{fig:d4000_rew},
our sample follows a similar overall trend, but generally
shows larger H$\alpha$+[N\,\textsc{ii}] equivalent widths
at a given $D_{\rm Balmer}$ than the sample of
\citet{kriek_h_2011}. Our galaxies also extend to higher
sSFRs in the right panel. These differences should be
considered alongside the lower stellar masses probed by
our sample. The overall trend is consistent with galaxies
moving from large equivalent width and weak break strength
toward small equivalent width and strong break strength
as recent star formation declines. This qualitative
consistency between our observations and the models from
\citet{kriek_h_2011} suggests that
H$\alpha$+[N\,\textsc{ii}] EW and $D_{\rm Balmer}$ together
provide a useful framework for tracing recent star
formation suppression in galaxies.

The right panel of Fig.~\ref{fig:d4000_rew} further supports this interpretation by comparing H$\alpha$+[N\,\textsc{ii}] equivalent width with specific SFR (sSFR). There is a general trend between sSFR and equivalent width, but the comparison with the \citet{kriek_h_2011} tracks is less direct in this plane. This difference likely arises from the dependence of sSFR on stellar mass, which varies considerably between our lower-mass sample and the more massive galaxies from \citet{kriek_h_2011}. Despite these mass differences, both samples follow a similar path in the $D_{\rm Balmer}$--EW plot, suggesting that $D_{\rm Balmer}$ is a relatively stable empirical indicator of recent star formation suppression across a range of stellar masses. Overall, the joint evolution of these two features in our sample supports recent or rapid suppression of star formation at cosmic noon.

\subsection{Comparison to Previous Results across Cosmic Time}

These conclusions are broadly consistent with findings
for post-starburst galaxies across cosmic time. In the local universe, post-starburst galaxies are characterized by strong Balmer absorption from intermediate-age stars and weak H$\alpha$ emission, indicating a recent end of star formation. These systems are typically linked to rapid quenching caused by gas-rich mergers or other processes \citep{wild_building_2009}. Their spectral properties imply a recent starburst followed by the truncation of star formation over tens to hundreds of Myr.

At higher redshifts ($z \sim 2$--$3$), post-starburst galaxies typically exhibit large continuum breaks and strong Balmer absorption, consistent with recent quenching on timescales of 300 to 800 Myr \citep{deugenio_hst_2021}. Spectroscopic surveys also show that quiescent galaxies at these redshifts often transitioned from star-forming to quenched within the past few hundred Myr \citep{deugenio_typical_2020}. Our results are consistent with this picture, because galaxies with larger $D_{\rm Balmer}$ values show weaker H$\alpha$+[N\,\textsc{ii}] emission and lower recent star formation activity.

Recent JWST investigations extend this trend to $z > 3$. \citet{carnall_massive_2023} confirmed the existence of a quiescent galaxy at $z = 4.658$ and inferred that its star formation had already been suppressed within the first 1.5 Gyr of cosmic history. Such discoveries show that strong suppression of star formation can occur early in the universe. Our sample connects this broader picture to a larger set of galaxies at $1<z<4$, where $D_{\rm Balmer}$ and recombination-line equivalent widths can be measured directly from JWST/NIRSpec spectra.

\subsection{Reliability of \texorpdfstring{$D_{\rm Balmer}$}{D Balmer} and Emission-Line Equivalent Widths as Evolutionary Tracers}

The $D_{\rm Balmer}$ index and H$\alpha$+[N\,\textsc{ii}] equivalent width are useful diagnostics of galaxy evolution, capable of tracing stellar population age and recent star formation activity. In this investigation, we use both indicators to characterize galaxies at cosmic noon, a critical epoch when many massive quiescent systems are being formed. Our results show that these diagnostics remain useful at these redshifts, while their interpretation requires attention to the low resolution of the PRISM spectra.

The $D_{\rm Balmer}$ index reflects the continuum contrast
across the 3646-\AA{} Balmer edge. Our observations, as
seen in Fig.~\ref{fig:ew_DBalmer}, show a clear
anti-correlation between $D_{\rm Balmer}$ and
H$\alpha$+[N\,\textsc{ii}] equivalent width. This behavior is consistent with aging stellar populations and agrees qualitatively with the synthetic star formation history tracks and low-redshift behavior seen in \citet{kriek_h_2011}.

Recent ALMA observations of spectroscopically confirmed ultramassive quiescent galaxies at $z>3$ also find weak or undetected dust-continuum emission in most targets, showing that massive quiescent galaxies at high redshift are not always dust-rich systems with hidden star formation \citep{chang2026magaz3ne}. This external evidence is consistent with the weak correlation between $D_{\rm Balmer}$ and SED-derived $E(B-V)$ in our sample, suggesting that the observed strong continuum breaks are driven mainly by evolved stellar populations instead of dust attenuation.

The H$\alpha$ equivalent width measures the ratio of current ionizing luminosity to the underlying continuum, so it is sensitive to very recent star formation. H$\alpha$ emission is powered by short-lived massive stars \citep{floresvelazquez_time-scales_2021}. Because these stars disappear quickly after massive star formation declines, the H$\alpha$ equivalent width also falls rapidly. The consistency of this trend with theoretical expectations and previous spectroscopic studies supports its use as a diagnostic at higher redshift.

The main limitation is that H$\alpha$ at 6563\,\AA\ is blended with [N\,\textsc{ii}] in low-resolution PRISM spectra. The measured equivalent width therefore includes both H$\alpha$ and [N\,\textsc{ii}], which can increase the inferred line strength, especially in metal-rich galaxies with higher [N\,\textsc{ii}]/H$\alpha$ ratios \citep{faisst_empirical_2018}. For this reason, we interpret the blended H$\alpha$+[N\,\textsc{ii}] EW as an empirical tracer of recent star formation activity, with possible contributions from metallicity, shocks, or AGN in some sources. In summary, $D_{\rm Balmer}$ and H$\alpha$+[N\,\textsc{ii}] equivalent width together are useful diagnostics of recent star formation suppression at cosmic noon.

\subsection{\texorpdfstring{$D_{\rm Balmer}$}{D Balmer} and the Star-Forming Main Sequence Offset}

In our sample, star formation rate correlates with stellar mass, as seen in Fig.~\ref{fig:residuals_D4000}, and galaxies with high $D_{\rm Balmer}$ systematically lie below the star-forming main sequence. In other words, galaxies with large $D_{\rm Balmer}$ indices form stars at a lower rate than main-sequence galaxies with the same stellar mass. In a recent JWST investigation, \citet{skarbinski_jwst_2025} found that quiescent galaxies at $z\sim2$--$3$ fall below the main sequence. This negative offset can be quantified as
\begin{equation}
\Delta \log{\rm SFR}_{\rm SED}
= \log({\rm SFR}_{\rm SED})
  - \log({\rm SFR}_{\rm MS}(M_\star)).
\end{equation}

Using this measure, galaxies with large $D_{\rm Balmer}$ indices have more negative offsets, as shown in Fig.~\ref{fig:residuals_D4000}. As a result, both $D_{\rm Balmer}$ and $\Delta \log{\rm SFR}_{\rm SED}$ track the same physical effect: as galaxies age and their recent star formation declines, their star formation rate drops below the value expected for main-sequence galaxies of the same mass.

This offset from the star-forming main sequence can also be used as a quenching diagnostic. Combined with the H$\alpha$+[N\,\textsc{ii}] EW trend, the main-sequence offset reinforces the interpretation that $D_{\rm Balmer}$ marks galaxies with suppressed recent star formation. The agreement between these independent diagnostics supports the use of $D_{\rm Balmer}$ as a practical spectroscopic tracer of galaxies transitioning away from the star-forming population at cosmic noon.

\section{Summary} \label{sec:summary}

We analyze JWST/NIRSpec PRISM spectra of galaxies at $z \sim 1$--$4$ to study how Balmer-break strength, $D_{\rm Balmer}$, relates to recent star formation activity and galaxy evolution at cosmic noon. Our main results are as follows.

\begin{itemize}
    \item $D_{\rm Balmer}$ correlates with mass-weighted stellar age from SED fitting. This supports the use of $D_{\rm Balmer}$ as an empirical tracer of evolved stellar populations in JWST/NIRSpec PRISM spectra at $1<z<4$.

    \item We find a strong anti-correlation between H$\alpha$+[N\,\textsc{ii}] equivalent width and the $D_{\rm Balmer}$ index. Galaxies with stronger Balmer breaks show weaker nebular emission, consistent with lower recent star formation activity and older stellar populations.

    \item Galaxies with stronger Balmer breaks tend to lie further below the star-forming main sequence at fixed stellar mass. This trend is consistent with the H$\alpha$+[N\,\textsc{ii}] EW result and indicates that $D_{\rm Balmer}$ is linked to recent suppression of star formation.

    \item We find little correlation between $D_{\rm Balmer}$ and SED-derived stellar $E(B-V)$. This indicates that dust attenuation is unlikely to be the main driver of the observed trends between $D_{\rm Balmer}$, emission-line equivalent width, and main-sequence offset.
\end{itemize}

These findings support the use of $D_{\rm Balmer}$ and recombination-line equivalent widths as practical diagnostics for identifying galaxies with suppressed recent star formation at cosmic noon. Future higher-resolution spectroscopy will help separate H$\alpha$ from [N\,\textsc{ii}], improve emission-line SFR measurements, and test how these trends depend on stellar mass, metallicity, and redshift.

\begin{acknowledgments}

We acknowledge support from the JWST program \#6541. Support for program \#6541 was provided by NASA through a grant from the Space Telescope Science Institute, which is operated by the Association of Universities for Research in Astronomy, Inc., under NASA contract NAS 5-03127.
YZ and CNAW acknowledge support by the JWST/NIRCam Science Team contract to the University of Arizona, NAS5-02105. S.C acknowledges support by European Union’s HE ERC Starting Grant No. 101040227 - WINGS.

This work is based on observations made with the NASA/ESA/CSA James Webb Space Telescope. The data were obtained from the Mikulski Archive for Space Telescopes at the Space Telescope Science Institute, which is operated by the Association of Universities for Research in Astronomy, Inc., under NASA contract NAS 5-03127 for JWST. These observations are associated with program \#6541. All of the data presented in this paper were obtained from the Mikulski Archive for Space Telescopes (MAST) at the Space Telescope Science Institute. The specific observations analyzed can be accessed via \dataset[https://doi.org/10.17909/habp-qe38]{https://doi.org/10.17909/habp-qe38}. STScI is operated by the Association of Universities for Research in Astronomy, Inc., under NASA contract NAS5-26555. Support to MAST for these data is provided by the NASA Office of Space Science via grant NAG5--7584 and by other grants and contracts.

We respectfully acknowledge the University of Arizona is on the land and territories of Indigenous peoples. Today, Arizona is home to 22 federally recognized tribes, with Tucson being home to the O'odham and the Yaqui. The university strives to build sustainable relationships with sovereign Native Nations and Indigenous communities through education offerings, partnerships, and community service.

This manuscript benefited from grammar checking and proofreading using ChatGPT \citep{openai_chatgpt_2024}. Part of the plotting code was generated by ChatGPT and ClaudeAI \citep{claude2024}.

\end{acknowledgments}

\begin{contribution}
S.N.\ did the analysis and wrote this paper. Y.Z.\ and E.E.\ contributed to the project design and results interpretation and discussion. S.C.\ reduced the spectra. F.D., C.dC. and CNAW proofread and commented on the paper.
\end{contribution}

\vspace{5mm}
\facilities{JWST, MAST}

\bibliography{Thesis3}{}
\bibliographystyle{aasjournalv7}

\end{document}